\documentstyle[12pt]{article}
\def\be{\begin{eqnarray}}
\def\ee{\end{eqnarray}}
\def\eq{\label}

\def\abstract#1{\vskip 7mm 
	\begin{center}{\large Abstract}\par \bigskip
		\begin{minipage}[c]{12cm}
			\small #1
		\end{minipage}
	\end{center}
}
\def\title#1{\begin{center}{\Large\bf #1}\end{center}}
\def\author#1{\vskip 5mm \begin{center}{#1}\end{center}}
\def\address#1{\begin{center}{\it #1}\end{center}}

\newcommand{\bfr}{\begin{flushright}}
\newcommand{\efr}{\end{flushright}}

\begin{document}

\vspace*{-0cm}
\title{Infinite number of soliton solutions to 5-dimensional vacuum Einstein equation}
\author{Takao KOIKAWA\footnote{E-mail: koikawa@nbilx02.nbi.dk \\On leave absence from Otsuma Women's University}}
\vspace{1cm}
\address{
        The Niels Bohr Institute,\\
         Blegdamsvej 17, 2100 Copenhagen, \\
         Denmark
}
\vspace{5cm}

\abstract{}
We give an infinite number of exact solutions to the 5-dimensional static Einstein equation with axial symmetry by using the inverse scattering method. The solutions are characterized by two integers representing the soliton numbers. The first non trivial example of these solutions is the static black ring solution found recently.

\newpage

In recent years interest in higher dimensional Einstein equation has increased. 
Especially 5-dimensional black ring solution discovered by Emaparaton and Reall\cite{ring, staticring, blackring} attracts considerable attention. In contrast to the 4-dimensional black hole case, 5-dimensional Einstein equations allow for more topologies besides $S^3$. The black ring solution, which has a regular event horizon characterized by $S^2 \times S^1$, is one of the interesting examples. The stationary black ring solution supported by the angular momentum is solved under the stationary and axi-symmetric conditions, while the static solution has a conical singularity.

In the 4-dimensional case, most of the well known exact solutions to the Einstein equation can be constructed as soliton solutions. The Einstein equations with some symmetries fall into completely integrable systems and so they have an infinite number of exact solutions. Most of the well known solutions are derived as the 2-soliton solutions in the series of multi-soliton solutions. The Kerr solution was first studied by using the inverse scattering technique in \cite{Bel-Zak}. We have studied the multi-Weyl solutions including the Schwartzschild solution in \cite{weyl}, the solutions to the Einstein equation with a scalar field in \cite{scalar}, multi-charged Weyl solution including the multi-Reissner Nordstr{\"o}m solution in \cite{R-N} and the dilaton black hole solutions with electric charge in \cite{dilaton}. These examples show that the inverse scattering is a powerful tool to study the Einstein equations in 4 dimension, and it is interesting to ask if we can also study the 5-dimensional Einstetin equations. 

We also studied the Kaluza-Klein type solution\cite{KK1,KK2,KK3}, where the extra-dimensions are assumed to be compactified like as tori. However, in this paper we do not distinguish the inner and the outer spaces and we do not impose such compactification conditions for extra dimensions. We study the 5-dimensional Einstein equations by use of the inverse scattering method. In order to solve the static vacuum solutions to the 5-dimensional Einstein equation we impose the axial symmetry and use the canonical variables $\rho$ and $z$. Though there can be some possible topologies in 5-dimensional case, we would like to find intriguing solutions including static ring solution as the first non-trivial solution.

We first derive the Einstein equations one of which has a matrix form. By assuming the static and axially symmetric conditions, we obtain general soliton solutions which are characterized by two soliton numbers and so we call it $(m, n)$-soliton solution using two integers $m$ and $n$. We then analyse the first few examples for smaller $m$ and $n$ and show that $(0,2)$-soliton solution is identical to the static ring solution using the axial symmetric variables $\rho$ and $z$. In the remaining we discuss the characters of the 5-dimensional black hole solutions, which might have more event horizons than two and may break the no hair theorem.


The metric is assumed to be given by
\be
ds^2=f(d\rho^2+dz^2)+g_{ab}dx^adx^b,
\ee
where $f$ and $g_{ab}$ are functions of $\rho$ and $z$ and the suffices $a$ and $b$ of $g_{ab}$ run over, 0, 1, 2. In the following we express the $3\times 3$ matrix by $g$. The Einstein equations in the above metric read
\be
&&\left(\rho g,_{\rho}g^{-1}\right),_{\rho}+\left(\rho g,_{z}g^{-1}\right),_{z}=0,\eq{eq:zero_curvature}\\
&&(\ln f),_{\rho}=-\frac{1}{\rho}+\frac{1}{4\rho}Tr(U^2-V^2),\\
&&(\ln f),_{z}=\frac{1}{2\rho}Tr(UV),
\ee
where $g,_\rho$ represents the partial derivative of $g$ with respect to the variable $\rho$, and $U$ and $V$ are $3\times 3$ matrices defined by
\be
U=\rho g,_{\rho}g^{-1},\quad V=\rho g,_{z}g^{-1}.
\ee
In the 4-dimensional case, the matrix equation (\ref{eq:zero_curvature}) has a $2 \times 2$ matrix form. In the higher dimensional Einstein equations it becomes a larger matrix form. In the study of the Kaluza-Klein general relativity\cite{KK2}, the matrix has a block-diagonal form composed of $2\times 2$ and $N\times N$matrices where $N$ represents the dimension of the extra space. The matrix form equation is also derived in the recent paper \cite{Troels}.  For the 5-dimensional case, it is the $3 \times 3$ matrix form. We assume the static condition in the present paper, and it implies that the matrix $g$ is a diagonal matrix. Noting that the matrix $g$ should satisfy the condition
\be
{\rm det} g=-\rho^2,
\ee
we define the matrix $g$ by
\be
g=\rm{diag}\left(-h_1^{-1}h_2^{-1},\left(\sqrt{\rho^2+z^2}-z\right)h_1,\left(\sqrt{\rho^2+z^2}+z \right) h_2 \right),
\ee
where $h_1$, $h_2$ are the functions of $\rho$ and $z$. The metric is explicitly written as
\be
ds^2&=&-h_1^{-1}h_2^{-1}dt^2+\left(\sqrt{\rho^2+z^2}-z\right) h_1 dx_1^2
+\left(\sqrt{\rho^2+z^2}+z \right) h_2 dx_2^2 \nonumber \\
&+&f(d\rho^2+dz^2).
\eq{eq:metric}
\ee
The matrix equation (\ref{eq:zero_curvature}) is rewritten to the equations for $h_i,~(i=1,2)$ as
\be
\left(\rho(\ln h_i),_{\rho}\right),_{\rho}+\left(\rho(\ln h_i),_{z}\right),_{z}=0.\quad (i=1,2)
\eq{eq:zero_curvature2}
\ee

In order to solve these equations, we first solve  (\ref{eq:zero_curvature2}) to obtain $g$, and then construct $U$ and $V$. By substituting the results to RHS of the above Einstein equations for $f$, we integrate them to obtain $f$. It has been known that the equation for $g$ is the zero-curvature equation\cite{Bel-Zak} just like that for the soliton equations. This is the reason why we call the solutions soliton solutions. In contrast to the 4-dimensional case, there are two soliton equations as in (\ref{eq:zero_curvature2}). Since each of them has the soliton solutions, we can specify the solution by their soliton integer numbers  $m$ and $n$. We shall call them $(m,n)$-soliton solution, which is explicitly given by
\be
h_1&=&\rho^{-m}\prod_{k=1}^m (i\mu_k),\\
h_2&=&\rho^{-n}\prod_{l=1}^n (i{\tilde \mu}_l).
\ee
Here $\mu_k$ and $\tilde \mu_l$ are called the poles given by
\be
\mu_k&=&w_k-z \pm \sqrt{\rho^2+(w_k-z)^2},\\
\tilde \mu_k&=&\tilde w_k-z \pm \sqrt{\rho^2+(\tilde w_k-z)^2},
\ee
where $w_k$ and $\tilde w_l$ are arbitrary constants. They can be either real or complex. Here we restrict ourselves to real poles. We suppress the deformation parameter here, because we are now interested in a series related to the static black ring solutions. By substituting $(m,n)$-soliton solution to $U$ and $V$, we integrate the above Einstein equations for $f$ to obtain
\be
f&=&C_{mn}\frac{\rho^{\frac{m^2+n^2-m-n+mn}{2}}}{\sqrt{\rho^2+z^2}}
\left|{\frac{\sqrt{\rho^2+z^2}-z}{\sqrt{\rho^2+z^2}+z}}\right|^{\frac{n-m}{4}} 
\left(\prod_{k=1}^m  \mu_k \right)^{-m-\frac{n}{2}+2}
\left(\prod_{l=1}^n {\tilde \mu}_k \right)^{-n-\frac{m}{2}+2} 
\cr
&\times&~
\prod_{k=1}^m \left(\sqrt{\rho^2+z^2}-z-\mu_k \right)
\prod_{l=1}^n \left(\sqrt{\rho^2+z^2}+z+{\tilde \mu}_l \right) \cr
&\times&~
\frac{
\prod_{k>l} \left(\mu_k-\mu_l \right)^2
\prod_{k>l} \left({\tilde \mu}_k-{\tilde \mu_l} \right)^2
\prod_{k} \prod_{l} \left( \mu_k-{\tilde \mu}_l \right)
}{
\prod_{k=1}^m \left(\mu_k^2+\rho^2 \right)
\prod_{l=1}^n \left(\tilde \mu_l^2+\rho^2 \right)
},
\eq{eq:f}
\ee
where $C_{mn}$ is a constant to be determined by the asymptotic flat condition.


We first consider the $(0,0)$-soliton solution, which is the flat solution. In this case, the metric components $h_1$, $h_2$ and $f$ are given by
\be
h_1=h_2=1,\\
f=\frac{1}{2\sqrt{\rho^2+z^2}},
\ee
where we have set $C_{00}=1/2$ in (\ref{eq:f}). We introduce the coordinates $r$ and $\theta$ by
\be
\rho=\frac{1}{2}r^2\sin 2\theta,\eq{eq:trans1}\\
z=\frac{1}{2}r^2\cos 2\theta.
\ee
We then notice that
\be
f&=&\frac{1}{r^2},\\
d\rho^2+dz^2&=&r^2(dr^2+r^2 d\theta^2).
\ee
Then, by writing the coordinates as $x^1=\phi$ and $x^2=\psi$, we obtain the flat metric
\be
ds^2=-dt^2+r^2 \sin^2 \theta d\phi^2+r^2 \cos^2 \theta d\psi^2+dr^2+r^2 d\theta^2.
\ee


Next we discuss the $(0,2)$-soliton solution, which is given by
\be
h_1&=&1,\\
h_2&=&-\frac{\tilde \mu_1 \tilde\mu_2}{\rho^2},\\
f&=&C_{02}\frac{\sqrt{\rho^2+z^2}-z}{\sqrt{\rho^2+z^2}}\left(\tilde\mu_2-\tilde\mu_1\right)^2 \nonumber \\
&\times &\frac{\left(\sqrt{\rho^2+z^2}+z+\tilde\mu_1\right)\left(\sqrt{\rho^2+z^2}+z+\tilde\mu_2\right)}{\left(\tilde\mu_1^2+\rho^2\right) \left(\tilde\mu_2^2+\rho^2\right)}.
\ee
We choose the minus sign for $\tilde\mu_1$ and the plus sign for $\tilde\mu_2$ and set $\tilde w_1=z_0-\sigma$ and $\tilde w_2=z_0+\sigma$ without loss of generality
\be
\tilde\mu_1&=&z_0-\sigma-z-\sqrt{\rho^2+(z+z_0+\sigma)^2},\\
\tilde\mu_2&=&z_0+\sigma-z+\sqrt{\rho^2+(z-z_0-\sigma)^2}.
\ee
Since we can always shift the origin in the $z$ axis, we are allowed to transform $z \to z+z_0$  in the metric. In addition to the transformed $\tilde \mu_i,~(i=1,2)$, we also define $\tilde \mu_3$ and their conjugates by
\be
\tilde\mu_1&=&-\sigma-z-R_1,\\
\tilde\mu_1^*&=&-\sigma-z+R_1,\\
\tilde\mu_2&=&\sigma-z+R_2,\\
\tilde\mu_2^*&=&\sigma-z-R_2,\\
\tilde\mu_3&=&-z_0-z-R_3,\\
\tilde\mu_3^*&=&-z_0-z+R_3,
\ee
where $R_i$ are given by
\be
R_1&=&\sqrt{\rho^2+(z+\sigma)^2},\\
R_2&=&\sqrt{\rho^2+(z-\sigma)^2},\\
R_3&=&\sqrt{\rho^2+(z+z_0)^2}.
\ee
They satisfy $\tilde \mu_i\tilde \mu_i^*=-\rho^2$ for i=1,~2,~3. The nonzero metric components $g_{ab}$ and $f$ are given by
\be
g_{00}&=&-\frac{\tilde \mu_1^*}{\tilde \mu_2}=-\frac{-\sigma-z+R_1}{\sigma-z+R_2},\\
g_{11}&=&\tilde \mu_3^*=-z_0-z+R_3,\\
g_{22}&=&\frac{\tilde \mu_3 \tilde \mu_2}{\tilde \mu_1^*}=\frac{(-z_0-z-R_3)(\sigma-z+R_2)}{-\sigma-z+R_1},\\
f&=&C_{02}\frac{\tilde \mu_3^* \left(\tilde\mu_2-\tilde\mu_1\right)^2 \left(-\tilde \mu_3+\tilde\mu_1\right)\left(-\tilde \mu_3+\tilde\mu_2\right)}{R_3\left(\tilde\mu_1^2+\rho^2\right) \left(\tilde\mu_2^2+\rho^2\right)}.
\ee 
We shall rewrite $f$ in order to compare the known solution. Using that
\be
\tilde \mu_1^2+\rho^2&=&-2\tilde \mu_1 R_1,\\
\tilde \mu_2^2+\rho^2&=&2\tilde \mu_2 R_2,
\ee
and the quadratic relations satisfied by $\tilde \mu_i$s and $\tilde \mu_i^*$s
\be
\mu_i^2-2(w_i-z)\mu_i-\rho^2=0,
\ee
we can rewrite $f$ as
\be
f=-\frac{C_{02}z_0}{2}\frac{\left(R_1+R_2+2\sigma \right)\left((1+\frac{\sigma}{z_0})R_1+(1-\frac{\sigma}{z_0})R_2-2\frac{\sigma}{z_0}R_3\right)}{R_1R_2R_3}.
\ee
This is identical to the result in \cite{Troels}.


We shall derive the 5-dimensional Schwarzshild solution as a special case of the static black ring solutions. In the above solution, the parameters $z_0$ and $\sigma$ are independent. But now we assume the relation $z_0=-\sigma$ between them. Then $R_3=R_2$, and so $f$ is simplified as
\be
f=2C_{02}\sigma\frac{R_1+R_2+2\sigma}{R_1R_2}.
\ee
By introducing $r$ and $\theta$ by
\be
\rho&=&\frac{1}{2}\left(1-\frac{4\sigma}{r^2}\right)^{\frac{1}{2}}r^2 \sin 2\theta,\eq{eq:trans3}\eq{eq:defofr}\\
z&=&\frac{1}{2}\left(1-\frac{2\sigma}{r^2}\right)r^2 \cos 2\theta,\eq{eq:defofth}
\ee
$R_i$ are written as
\be
R_1&=&\frac{r^2}{2}-2\sigma \sin^2 \theta,\\
R_2&=&R_3=\frac{r^2}{2}-2\sigma \cos^2 \theta,
\ee
and so $\tilde\mu_i$ (i=1,2,3) and their conjugates are expressed as
\be
\tilde\mu_1&=&-r^2 \cos^2 \theta,\\
\tilde\mu_1^*&=&(r^2-4\sigma) \sin^2 \theta,\\
\tilde\mu_2&=&\tilde \mu_3^*=r^2 \sin^2 \theta,\\
\tilde\mu_2^*&=&\tilde \mu_3=-(r^2-4\sigma) \cos^2 \theta.
\ee
Then $g_{ab}$ and $f$ are to be rewritten as
\be
g_{00}&=&-\left(1-\frac{4\sigma}{r^2}\right),\\
g_{11}&=&r^2 \sin^2 \theta,\\
g_{00}&=&r^2 \cos^2 \theta,\\
f&=&\frac{8C_{02}\sigma}{r^2\left(1-\frac{4\sigma}{r^2}+\frac{4\sigma^2\sin^2 2\theta}{r^2}\right)}.
\ee
We also note that
\be
d\rho^2+dz^2 &=&r^2\left(1-\frac{4\sigma}{r^2}+\frac{4\sigma^2\sin^2 2\theta}{r^2} \right)
\left( \left(1-\frac{4\sigma}{r^2}\right)^{-1}dr^2 +r^2d\theta^2 \right).\nonumber \\
\ee
Setting the constant $C_{02}=\frac{1}{8\sigma}$, we obtain the 5-dimensional Schwartzshild  solution as a special case of the $(0,2)$-soliton soliton solution:
\be
ds^2=-(1-\frac{4\sigma}{r^2})dt^2+(1-\frac{4\sigma}{r^2})^{-1}dr^2+r^2 d\theta^2+r^2 \sin^2 \theta d\phi^2+r^2 \cos^2 \theta d\psi^2.
\ee

We next consider the $(2,0)$-soliton solution. The metric components $h_1$, $h_2$ and $f$ are given by
\be
h_1&=&-\frac{\mu_1\mu_2}{\rho^2},\\
h_2&=&1,\\
f&=&C_{20}\frac{\sqrt{\rho^2+z^2}+z}{\sqrt{\rho^2+z^2}}\left(\mu_2-\mu_1\right)^2\\
&\times &\frac{\left(\sqrt{\rho^2+z^2}-z-\mu_1\right)\left(\sqrt{\rho^2+z^2}-z-\mu_2\right)}{\left(\mu_1^2+\rho^2\right) \left(\mu_2^2+\rho^2\right)}.
\ee
Here we choose the minus sign for $\mu_1$ and the plus sign for $\mu_2$ and set 
$w_1=z_0-\sigma$ and $w_2=z_0+\sigma$ as in the previous assignments for $\tilde \mu_1$ and $\tilde \mu_2$
\be
\mu_1&=&z_0-\sigma-z-\sqrt{\rho^2+(z-z_0+\sigma)^2},\\
\mu_2&=&z_0+\sigma-z+\sqrt{\rho^2+(z-z_0-\sigma)^2}.
\ee
We may translate in the $z$ direction $z\to z+z_0$ as before. Because of the slight difference from the $(0,2)$-soliton case in $g_{11}$ and $g_{22}$ as well as in $f$, the solution seems to be different from the $(0,2)$-soliton solution. However, in the special case that $z_0=\sigma$, we obtain the same 5-dimensional Schwarzschild metric by expressing the metric in terms of $r$ and $\theta$ defined in (\ref{eq:defofr}) and (\ref{eq:defofth}).


We have used the canonical coordinates $\rho$ and $z$ in solving the 5-dimensional Einstein equations, which have been extensively used in the study of the 4-dimensional solutions in the previous works. These can be found for the Kerr solution in \cite{Bel-Zak}, the multi-Weyl solutions including the Schwartzschild solution in \cite{weyl}, the solutions to the Einstein equation with a scalar field in \cite{scalar}, multi-charged Weyl solution including the multi-Reissner Nordstr{\"o}m solution in \cite{R-N} and the dilaton black hole solutions with electric charge in \cite{dilaton}. In these papers, the zeros of $\rho$ are either the positions of the horizons or the singularities of the black hole solution. This is related to the fact that ${\rm det}g=-\rho^2$. We have to note that the inverse is not always the case since ${\rm det}g$ is composed of the products of the metric components and so the zeros might be cancelled in the products. We shall list them up in the below.

In the case of the Kerr solution, the transformation of the canonical variables  $\rho$ and $z$ to the spherical coordinates $r$ and $\theta$ is given by\cite{Bel-Zak}
\be
\rho=\sqrt{(r-m)^2-\sigma^2}\sin \theta.
\ee
Here $\sigma$ is given by
\be
\sigma^2=m^2-a^2,
\ee
where $m$ is the mass of the black hole and $a$ is the angular momentum per unit mass. The zeros of $\rho$ at $r=m\pm \sqrt{m^2-a^2}$ correspond to the inner and outer horizons of the Kerr black hole solution. In the $a=0$ case, we obtain the Schwartzschild solution which is also obtained by setting the distortion parameter to be identity in the Weyl solution. In this case the above relation reduces to
\be
\rho=\sqrt{(r-2m)r}\sin \theta.
\ee
The zeros of $\rho$ in this case correspond to the horizon at $r=2m$ and the singularity at $r=0$.

In the Reisner-Nordstr{\"o}m solution, the metric component $g_{00}$ is given by
\be
g_{00}=1-\frac{2m}{r}+\frac{e^2}{r^2},
\eq{eq:RN}
\ee
where $m$ is the mass and $e$ is the electric charge. The solution can be also regarded as the 2-soliton solution of an infinite series of soliton solutions obtained by the inverse scattering method by adopting the axially symmetric coordinates $\rho$ and $z$\cite{R-N}.  We change the coordinates to the spherical coordinates $r$ and $\theta$, and  the variable $\rho$ is expressed as
\be
\rho=\sqrt{(r-m)^2-d^2}\sin \theta,
\ee
with
\be
d^2=m^2-e^2.
\ee
The zeros of $\rho$ are at $r=m\pm \sqrt{m^2-e^2}$, which are the positions of the inner and the outer horizons as is also seen from (\ref{eq:RN}).
 
In the dilaton black hole solution\cite{Gib-Mae}, the metric is given by
\be
ds^2=-(1-\frac{2M}{r})dt^2+(1-\frac{2M}{r})^{-1}dr^2+r(r+2\Sigma)d\Omega^2,
\ee
where $M$ is the mass of the black hole and $\Sigma$ is the dilaton charge. In order to prevent the area measured at some $r$ from vanishing and so getting singular as $r \to -2\Sigma$, the mass and the dilaton charge should satisfy the condition
\be
M>-\Sigma.
\ee
We can derive an infinite number of solutions including the above solution by the inverse scattering method. We use the spherical coordinates $r$ and $\theta$ from $z$ and $\rho$ to derive the above metric. The variable $\rho$ is related to them by 
\be
\rho=\sqrt{(r-2M)(r+2\Sigma)}\sin \theta.
\ee
This shows that the zeros of $\rho$ are the horizon at $r=2M$ and the singularity at $r=-2\Sigma$.

These examples in 4 dimensions show that the zeros of $\rho^2$ obtained by solving the quadratic equation of the spherical coordinates $r$ are either the horizons or singularities. Consider the general quadratic equation given by
\be
r^2+c_1r+c_0=0,
\ee
where the coefficients $c_0$ and $c_1$ are some constants. In order that the quadratic equation have two positive roots, the necessary condition is that both $c_0$ and $c_1$ should not be nonzero, in which two constants correspond to two charges of black holes. When $c_0=0$ and $c_1 \ne 0$, the roots are $r=0$ and $r=-c_1$. Here the nonzero constant corresponds to only hair or the mass of the black hole. This shows that we see the number of horizons and the charges from that of the roots of the quadratic equation.

Also in the 5-dimensional case, we use the canonical coordinates $\rho$ and $z$, but they are different from the 4-dimensional case. While $\rho$ has the dimension of length in 4-dimensional case, it has the dimension of $({\rm length})^2$ in the 5-dimensional case. The variable $\rho$ is related to the spherical coordinates $r$ and $\theta$ as in Eqs.(\ref{eq:trans1}) and (\ref{eq:trans3}). The zeros of $\rho^2$ is obtained by solving the quartic equation in contrast to the quadratic equations in the 4-dimensional case
\be
r^4+c_3r^3+c_2r^2+c_1r+c_0=0.
\ee
In the analogous discussion to the 4-dimensional case, this quartic equation  seems to suggest that the 5-dimensional black holes might have maximally four event horizons and four charges. In the vacuum case, the black hole can have only mass parameter, which is the result from $c_2 \ne 0$ while $c_3=c_1=c_0=0$. More constants would become nonzero and give rise more roots to the quartic equation if the stationary black hole solutions of nonvacuum Einstein equations with some fields using the canonical variables are obtained. We may incorporate the fields like the electro-magnetic fields, three-form field strength and the dilaton fields to solve the Einstein equations, and then we would have nonzero coefficients in the quartic equation. The black holes with those charges would show different features from the 4-dimensional case.


In this paper we obtain an infinite number of soliton solutions to the 5-dimensional vacuum Einstein equation by assuming the static and axially symmetric conditions, which is compactly expressed in terms poles. The solution is characterized by two soliton numbers, and the first non-trivial solution is the static black ring solution found recently. We can also apply the inverse scattering method to the stationary case, where we may find the singularity-free black hole solutions and so the study of the multi-soliton solutions would be more interesting. It may have precedence over studying the static case. It is also interesting to study the non-vacuum case by using the canonical variables, where we might find more event horizons than two and the more charges than three in contrast to the 4-dimensional black holes.

\end{document}